\documentclass[cits]{PoS}

\title{Radio Emission from Exoplanets}

\ShortTitle{Radio emission from exoplanets}

\author{\speaker{Samuel J. George}%
        \\
        School of Physics and Astronomy, University of Birmingham, UK\\
        E-mail: \email{samuel@star.sr.bham.ac.uk}}

\author{Ian R. Stevens\\
       School of Physics and Astronomy, University of Birmingham, UK\\
        E-mail: \email{irs@star.sr.bham.ac.uk}}

\abstract{We present results from new low frequency observations of two extrasolar planetary systems ($\epsilon$ Eridani and HD128311) taken at 150 MHz with the Giant Metrewave Radio Telescope (GMRT). We do not detect either system, but are able to place tight upper limits on their low frequency radio emission.}

\FullConference{From Planets to Dark Energy: the Modern Radio Universe\\
                 October 1-5 2007\\
                 The University of Manchester, UK}

\begin{document}

\section{Introduction}
Low frequency radio emission from solar system planets has been observed for decades and Jupiter (for example) is an extremely bright (though variable) source at decametric wavelengths. Below a cut-off frequency of $\sim 40$~MHz, the Jovian radio emission is thought to be dominated by cyclotron-maser processes from keV electrons in the auroral regions of the planet. We expect that, like Jupiter, exoplanets will have a dynamo driven magnetic field and a similar interaction with their host star. If this is the case then they will produce radio emission.  The major difference between these exoplanets and the planets in our own solar system will be the sensitivity that is required and the frequencies to undertake any observations (Stevens 2005, Zarka et al. 2001). Observations of planets at radio wavelengths have the potential for determining planetary properties, such as the rotation period and magnetic field strength, that are important in understanding the planetary structure.

Based on our knowledge of the solar system, the level of radio emission from a given magnetized exoplanet will be proportional to the stellar-wind ram-pressure flux incident on the planetary magnetosphere (Lazio et al. 2004; Stevens 2005). This means that the expected level of radio emission from a given exoplanet will be a function of several factors including; the planetary magnetic moment (scaling roughly with the planetary mass and the moment of inertia), the separation between the planet and the star,  the stellar wind properties of the stars and, of course, the radio power received scales with distance from the source. The optimum situation for detection would be a nearby system with a high mass planet in a short period orbit.
A number of previous searches have tended to focus on short period exoplanets, which from simple scaling laws are expected to be the most luminous. In contrast, we chose to investigate longer period systems, since exoplanets which are particularly close to their host star might experience tidal locking which could cause a negligible dynamo action. Using the criteria discussed above we decided to observe 2 known exo-planetary systems: $\epsilon$ Eridani, HD~128311 - both offer longer period planets than had been previously observed.

\section{Observations and Results}
\begin{figure*}
\begin{center}
\includegraphics[scale=0.45]{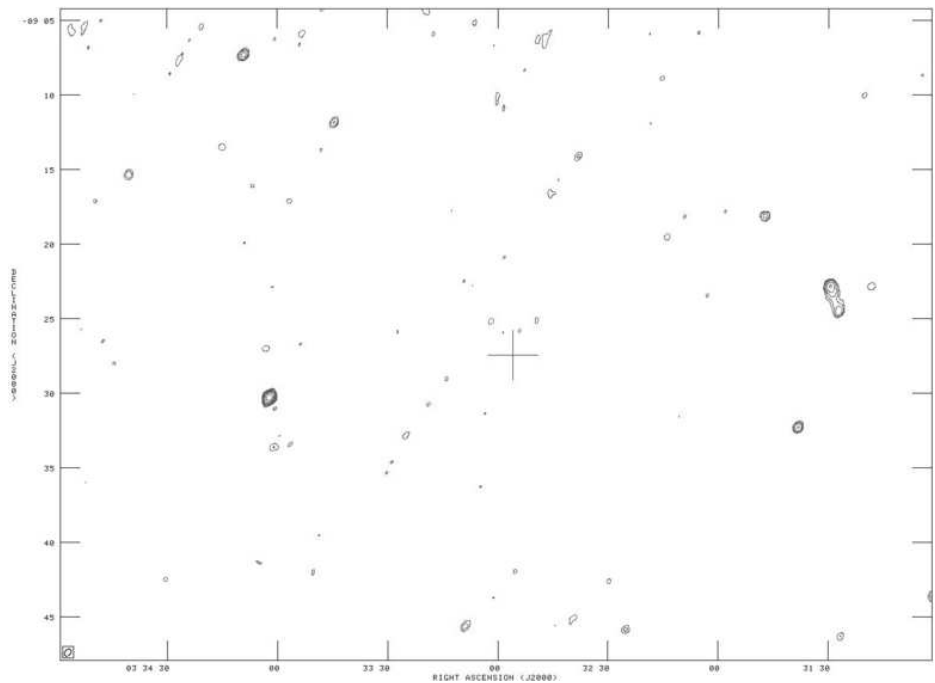}
\caption{The 150~MHz view of the region around $\epsilon$ Eri
(J2000.0 position marked with a cross). The contours levels are 12.5, 25, 50, 100, 200, 400, 800 and 
1600 mJy.}
\label{eps_eri}
\end{center}
\end{figure*}

The observations of $\epsilon$ Eri and HD~128311 were conducted with a central frequency of 157.0 MHz and a bandwidth of 8~MHz. Time on source in a single scan was 30 minutes followed by 5 minutes of phase calibration.
$\epsilon$ Eri and HD~128311 were observed for 4.13 hours each. 
The data was calibrated using standard procedures in AIPS. Unfortunately, at low frequencies observations are hindered by significantly RFI. This was identified and excised manually using AIPS procedures. Wide field mapping with 5 degree diameter (close to full width of the beam) consisting of 121 facets were performed to correct for curvature of the sky. Several iterations of phase-only self calibration was applied to correct for short term phase changes. After phase-only self calibration has converged, one round of amplitude and phase self calibration was carried out. A map of the area around $\epsilon$ Eri can be seen in Figure \ref{eps_eri}. The final restoring beam is $32.1'' \times 23.2''$ for $\epsilon$  Eri and $45.4'' \times 30.6''$ for HD128311. For the two observations the RMS noise levels are $\sigma = 3.16$~mJy ($\epsilon$  Eri) and $\sigma = 6.20$~mJy (HD128311), these tight constraints are comparable to the theoretical expected values. For a detailed discussion the reader is directed to George \& Stevens (2007).

\section{Background Sources and Future Observations}
In the full field of view (of area $\sim 10^\circ \times 10^\circ$) for both the $\epsilon$  Eri and HD128311 fields we have detected a large number of background sources. We are currently investigating the nature of these sources. We believe that they are most likely background AGN but at 150~MHz one might also expect to detect the presence of a starburst galaxy population (they will have a steeper spectrum). 

From our observations it is clear that a detection of an exoplanet at the frequencies considered is not practical, even with the most sensitive of instruments. There is great hope that LOFAR (operating at between 30-250MHz) might well enable the detection of emission from a number of exoplanets.


\begin{thebibliography}{99}
\bibitem{Dulk}G.A.~Dulk, et al., \emph{Search for Cyclotron-maser Radio Emission from Extrasolar Planets}, 1997, \emph{Bulletin of the American Astronomical Society}, 29, 2803
\bibitem{George}S.J.~George, I.R.~Stevens, \emph{GMRT Low Frequency Observations of Extrasolar Planetary Systems}, 2007 \emph{MNRAS}, 382, 455 {\tt arXiv/0708.4079}
\bibitem{Lazio}T.J.~Lazio, et al., \emph{The Radiometric Bode's Law and Extrasolar Planets}, 2004, \emph{ApJ}, 612, 511
\bibitem{Stevens}I.R.~Stevens, \emph{Magnetospheric radio emission from extrasolar giant planets: the role of the host stars}, 2005, \emph{MNRAS}, 356, 1053
\bibitem{Zarka}P.~Zarka, et al., \emph{Magnetically-Driven Planetary Radio Emissions and Application to Extrasolar Planets}, 2001, \emph{Astrophysics and Space Science}, 277, 293

\end{thebibliography}
\end{document}